# Bayesian Parameter Estimation for Predictive Modeling of Illumination-Dependent Current-Voltage Curves

Eunchi Kim[a], Thomas Kirchartz[a,b*]
[a] *IMD3-Photovoltaik, Forschungszentrum Jülich, 52425 Jülich, Germany*
[b] *Faculty of Electrical Engineering and Information Technology, RWTH Aachen, 52062 Aachen, Germany*

Machine learning enables rapid estimation of material parameters in solar cells via neural-network-based surrogate models. However, since multiple parameter combinations can reproduce the identical experimental data, it is often difficult to judge from the fitted data alone whether inferred parameters represent true physical properties. In this work, we argue that predictive power, which is the ability of the inferred parameter set to successfully describe the physical reality, provides a quantitative criterion for evaluating parameter estimation models. We demonstrate this framework using light-intensity-dependent current-voltage (*JV*) characteristics of organic solar cells by varying the treatment of dark shunt resistance and the fitting objective (raw $J$ vs. shifted current $J + J_{sc}$). Our analysis reveals that predictive performances under illumination conditions beyond the fitting data depend on parameter estimation models. These prediction results provide insights into which physical mechanisms are captured by each model. Moreover, we show that the predictive performance is also limited by the information content of the input data. In particular, reliable parameter estimation requires a dark *JV* curve and at least one illuminated *JV* curve, preferably measured at a light intensity where series resistance effects are not dominant. Together, these results demonstrate that predictive performance provides a practical and general framework for evaluating parameter estimation models.

## I. INTRODUCTION.

As interest in self-driving laboratories continues to grow for accelerating material discovery [1–10] and device optimization [11–17], the accurate identification of material parameters becomes increasingly important. These parameters can serve as essential inputs for feedback algorithms that guide the next experimental steps in autonomous research systems. Often, the accurate quantification of material parameters from experimental observables requires the inversion of a mathematical model that is used to approximate the physical reality of the experiment. For solar cell devices, this often involves drift-diffusion simulations, which require numerical solutions of coupled differential equations. Here, such an inversion may often be impossible to do analytically. Instead, the comparison of the mathematical model with the experimental observable is often only possible by fitting the data using regression algorithms that solve the forward problem of simulating the experiment often hundreds or thousands of times until they achieve good agreement. A traditional approach to solving this inversion problem is to evaluate fitting results for uniformly sampled points [Fig. 1(a)]. While it is the most straightforward approach to implement, the result is strictly limited to sampling density. If the grid is not sufficiently dense, the true global optimum may be missed. Moreover, this limitation becomes more severe as the number of varied parameters increases due to the exponential increase in computational cost.

To address this challenge, more efficient optimization strategies can be employed. One approach is to fit experimental data with a drift-diffusion solver using a Bayesian Optimization (BO) algorithm [Fig. 1(b)] [18–20]. This method has high flexibility in the choice of fitting parameters, allowing parameters to be added or removed as needed if the fit is bad. However, the BO model can become computationally demanding with extended evaluation histories and its performance is sensitive to the choice of hyperparameters. An alternative approach is to construct neural-network (NN)-based surrogate models that can replace the mathematical model [Fig. 1(c)], as initially proposed in the context of photovoltaic research by Ren et al [21]. Although the initial step of training an NN model can be time-intensive (1-2 days), this method ensures faster computation speed compared to a drift-diffusion solver. This, in turn, allows the use of simpler optimization methods such as genetic algorithms for parameter extraction. It should be noted that the effectiveness of such surrogate models depends on the availability of training data and is most suitable for systems with similar underlying physics. Both approaches enable exploration of high-dimensional parameter spaces but also provide estimates of the probability distribution for material parameters. In this context, the both processes can be categorized to Bayesian parameter estimation, where the probability is calculated given the observed evidence (experimental data).

*Contact author: t.kirchartz@fz-juelich.de

**(a) Deterministic / DD**

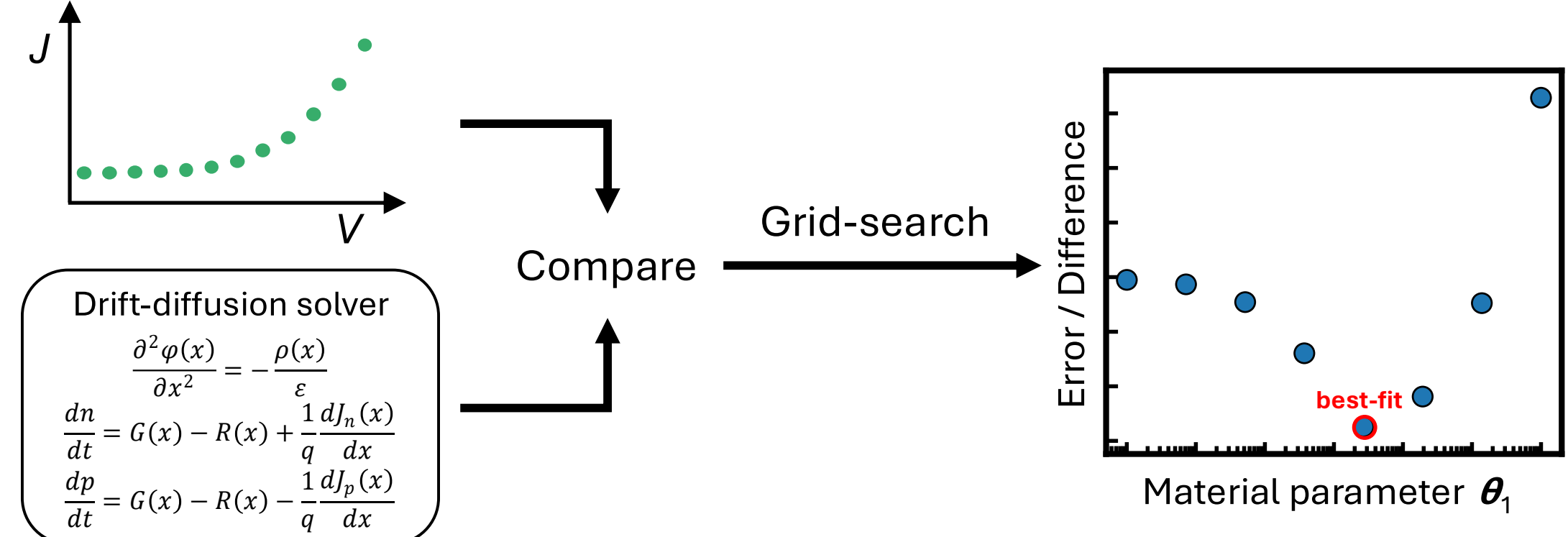

**(b) Bayesian / DD** (flexible)

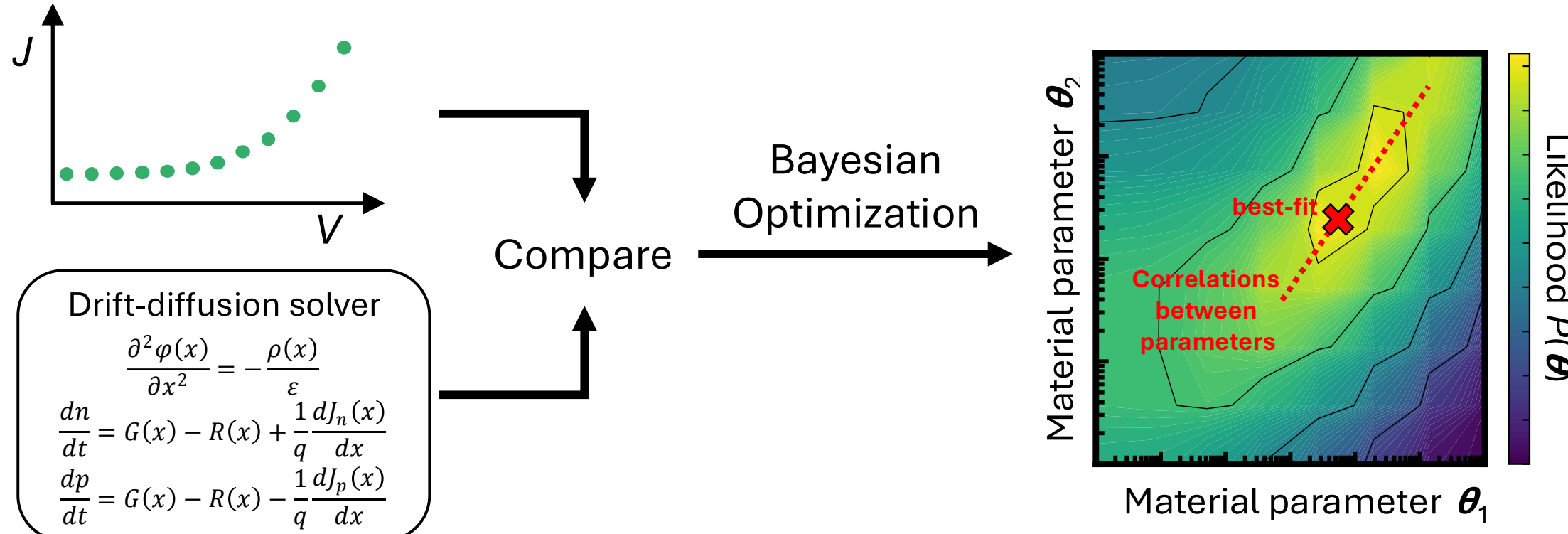

**(c) Bayesian / NN-based surrogate model** (fast, useful for many similar samples)

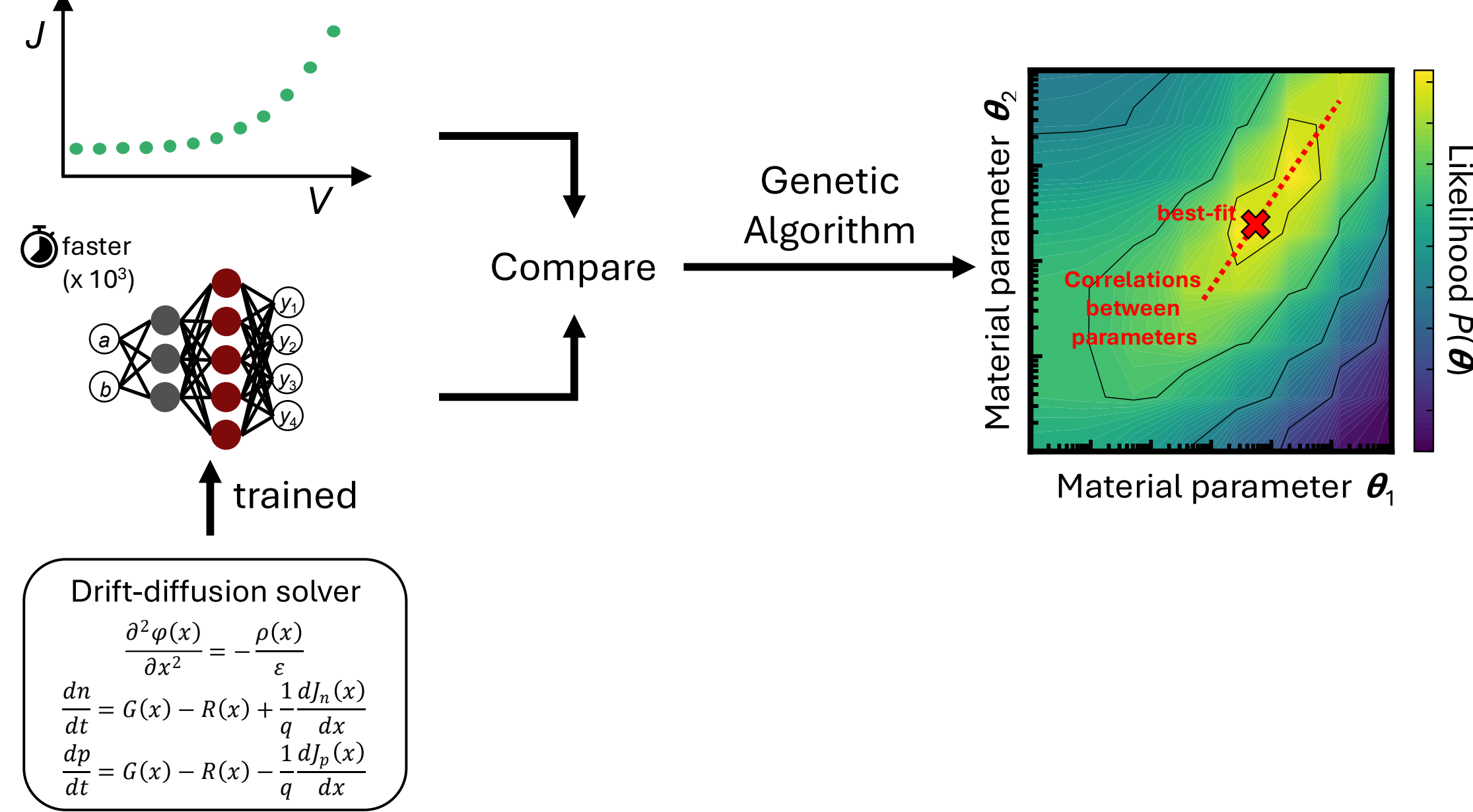

FIG 1. Schematic illustrations of three parameter estimation approaches. Each strategy identifies best-fit parameters $\theta_{\text{best-fit}}$ by evaluating the discrepancy between experimental current-voltage ($JV$) curve and results from drift-diffusion (DD) simulation or a trained surrogate model. (a) A conventional approach that evaluates uniformly sampled points and identifies $\theta_{\text{best-fit}}$ as the minimum error point. Although it is the simplest to implement, its accuracy is limited to the sampling density, making it difficult to guarantee

the global optimum. Furthermore, this method faces the "curse of dimensionality" in which the computational effort increases exponentially when the number of varied parameters increases. (b) An iterative approach using the DD solver directly to search for $\theta_{\text{best-fit}}$. Bayesian optimization (BO) offers high flexibility in selection of parameters of interest without prior model training. However, its computational cost scales poorly with longer evaluation histories and its performance depends highly on hyperparameter configurations. (c) A two-step approach beginning with the construction of a neural network (NN) model trained on a DD simulation dataset. While initial model construction is time-intensive (1-2 days), the resulting surrogate model ensures rapid and robust computation for high-throughput analysis. This facilitates the use of genetic algorithms which are simpler to implement than BO due to having fewer hyperparameters and no acquisition functions. Both (b) and (c) enable estimation of the probability distribution for material parameters, providing additional insights to parameter correlations and uncertainty.

A decisive yet inevitable step in parameter extraction is the definition of those material parameters that are allowed to vary to reproduce the experimental data. Those parameters will be the ones that the researcher conducting the study would classify as belonging to a class of parameters that either has a very high uncertainty (cannot independently be measured) and/or has a significant impact on the result. Parameters that are either well known or largely irrelevant would be excluded to reduce the dimensionality of the problem. Although this reduction is essential for computational tractability and for obtaining interpretable results, the choice of inferred parameters remains one of the least systematic parts of the process and often relies heavily on the experience of the researchers. In the so far quite limited literature on Bayesian inference in the context of photovoltaics, the initial step of determining which parameters to treat as free (high significance) and which to fix (low significance) has typically been approached in two ways. A common strategy is to incorporate insights from additional characterization techniques. For example, Hawks et al. [22] conducted photocurrent spectroscopy to understand the characteristics of defect states and included this information when fitting temperature-dependent current-voltage curves with a drift-diffusion model. Similarly, Tang et al. [23] utilized various characterization tools to identify the cause of s-shaped current-voltage curves and included the built-in voltage as one of the free parameters to reproduce experimental data. Alternatively, some groups [24–26] relied on an extensive literature search on the material system of interest and fixed certain material parameters based on previously reported values. While they leverage valuable prior physical knowledge to reduce uncertainty in the inference problem and improve parameter identifiability, the rationale behind parameter selection is not explicitly described or is only partly explained (TABLE S1 in Supplemental Material [47]). This, therefore, motivates the need for establishing more objective metrics to evaluate parameter estimation models.

Evaluating such models requires distinguishing between extracting the true physical properties and obtaining a useful physical description of the device. In particular, in the case of solving an inverse problem, multiple parameter combinations can produce the same experimental observation, making it difficult to determine the correctness of the inferred parameters from experimental data alone. Consequently, the value of a parameter estimation model should not be solely judged by whether the extracted parameter values represent the true physical quantities, but by whether the combination of the model, its underlying assumptions, and the inferred parameters provides a useful description of physical reality. In this context, the term "useful" refers to whether the combination of model and model-parameters has predictive power or if it can explain observed trends. Predictive power is particularly compelling as a criterion as it implies that the already known data is successfully described and in addition new data that the algorithm has not seen can be predicted. Therefore, in this work, we consider the usefulness of a model and its parameters (i.e. have predictive power) rather than the correctness of model parameters. This is primarily a pragmatic strategy, because predictive power can be quantified while correctness is nearly impossible to judge, when looking at a limited set of experimental data.

Previously, we developed an approach based on a NN model to extract material parameters $\theta$ such as mobilities and recombination coefficients from light-intensity-varied current-voltage (*JV*) curves [27]. With the constructed NN model at hand, it is possible to infer best-fit material parameters $\theta_{\text{best-fit}}$ from characterization data. However, there are numerous ways to construct this NN model or, more generally, a parameter estimation model and allowing all possible parameters to vary freely would, in principle, be expected to yield the most accurate model. In practice, however, it is difficult to make such a model, since it quickly becomes computationally intractable and weakens the interpretability of the inferred parameters. Therefore, our task is to find a useful model with a medium complexity, and, accordingly, in this manuscript we investigate how well a given, finitely parametrized model can predict device behavior under a defined set of boundary conditions. Additionally, we aim to understand which physical assumptions in the estimation model are missing or included that it leads

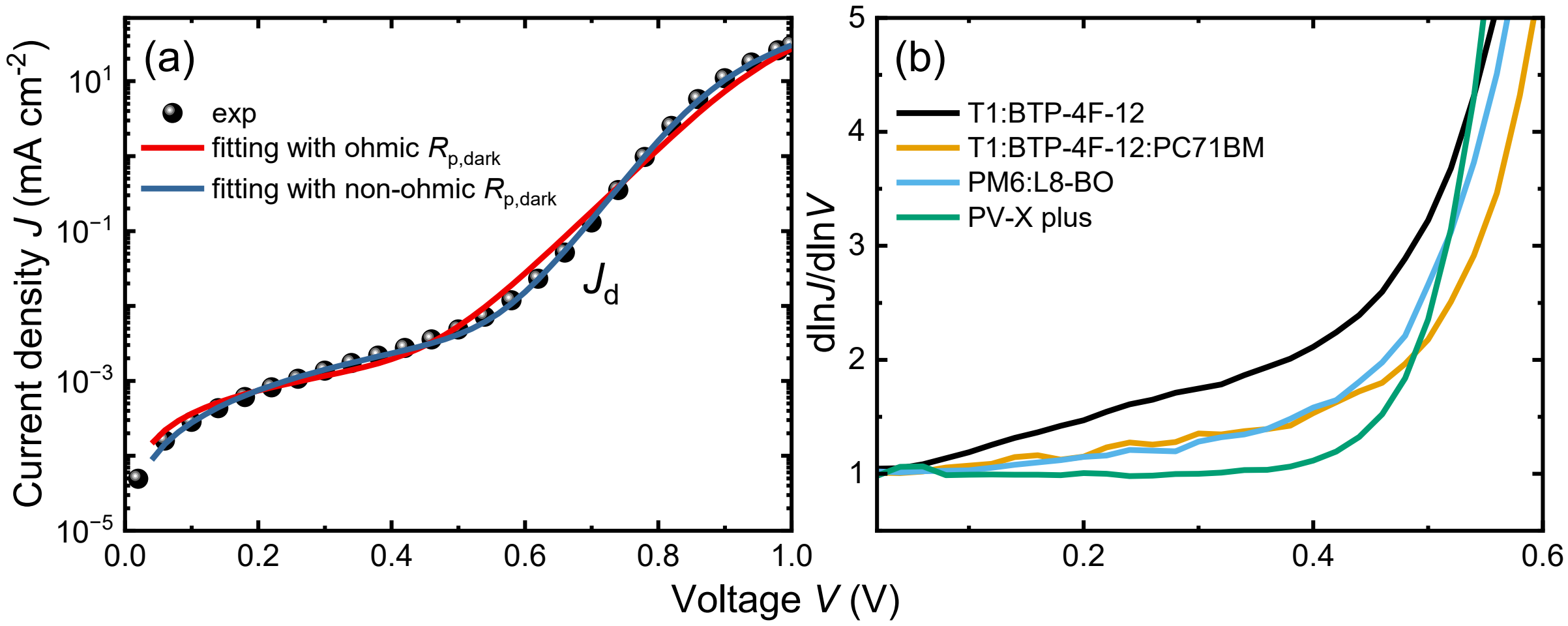

FIG 2. (a) Dark current-voltage curve of experimental data (dots) fitted using an equivalent circuit model with ohmic (red solid line) and non-ohmic (blue solid line) dark shunt resistance $R_{p,dark}$. (b) $d\ln J/d\ln V$ plotted against voltage for various absorber layer materials – T1:BTP-4F-12, T1:BTP-4F-12:PC71BM, PM6:L8-BO, and PV-X plus. The value of 1 indicates a linear relationship between current and voltage, while deviation from 1 suggests non-linear behavior.

to differences in predictive power for illumination-dependent device performance.

## II. RESULTS AND DISCUSSION

### A. Physical considerations for parameter estimation model

The predictive performance of the inferred parameter set fundamentally depends on whether the underlying model captures the dominant physical mechanisms governing the light-intensity dependence of the *JV* characteristics. The following discussion identifies these mechanisms and motivates the modifications in the parameter estimation workflow. In the low-light regime, the device performance is typically limited by defect-assisted recombination. However, when the dark shunt resistance $R_{p,dark}$ is sufficiently low, it becomes a dominant factor in determining efficiency [ [28–31]]. It becomes the critical point for recombination losses that do not further decrease with reduced light intensity. Therefore, it is important to model the $R_{p,dark}$ accurately when fitting the illuminated *JV* curves. Conventionally, $R_{p,dark}$ is considered an ohmic resistance [32–35]. In practice, however, it is often the case that the experimentally obtained dark *JV*s cannot be reconstructed with a linearly voltage-dependent $R_{p,dark}$ [36–40]. The nonlinear behavior of shunt resistance typically arises from the presence of pinholes in the absorber layer, providing an alternative current path between the transport layers [37]. Figure 2 supports these findings, showing that incorporating a non-ohmic $R_{p,dark}$ results in a more accurate fit to the experimental data. Furthermore, Figure 2(b) illustrates the power coefficient $\alpha$ in the relation $J_d \propto V^{\alpha}$, where $\alpha > 1$ indicates non-ohmic behavior. Most of the material systems exhibit higher $\alpha$ values, suggesting a nonlinear shunt behavior. Furthermore, for some active layer materials, we observe a thickness-dependence of the linearity of the shunt resistance. Specifically, we observe that $R_{p,dark}$ typically becomes more ohmic with increasing thickness (Fig. S1 in Supplementary Material [47]). Thus, verifying the presence of a non-ohmic $R_{p,dark}$ is important for accurate predictions of device performance in thin-film solar cells, particularly under low-light conditions. As light intensity increases, the influence of defect-assisted or bimolecular recombination becomes more significant. The type of the dominant recombination mechanism determines the slope of the open-circuit voltage $V_{oc}$ as a function of light intensity. For example, if the recombination is dominated by mid-gap trap states, the ideality factor $n_{id}$ increases, leading to a steeper drop in $V_{oc}$ with increasing light intensity, as described by $dV_{oc}/d\ln\phi \propto n_{id}$. At higher light intensity ($\phi$ ~ 1 suns), series resistance predominantly reduces fill factor, causing voltage losses near $V_{oc}$. Unlike shunt resistance, incorporating series resistance into fitting procedures is challenging. Therefore, capturing $n_{id}$ and the associated recombination coefficients or lifetime parameters is a key factor in understanding and predicting the light-intensity-dependent power conversion efficiency (PCE), particularly at high irradiance. In this regard, analyzing the *JV* curves in terms of shifted current density $J_{shifted} = J + J_{sc}$ can be especially useful to gain deeper insight into recombination dynamics. This shifted current is directly related to the recombination current $J_{rec}$ ($J_{shifted}(V, \phi) = J_{rec}(V, \phi)$ - $J_{rec}(V{=}0, \phi)$) and therefore encapsulates key information about recombination behavior, including the ideality factor and the mobility-lifetime product [29,41–43]. For example, $J_{shifted}$ exhibits a weak or near-linear dependence on the voltage at small forward bias, where this effect is referred to as apparent shunt resistance $R_{p,photo}$. Unlike the dark shunt resistance, the $R_{p,photo}$ originates from nonzero charge carrier density or high quasi-Fermi-level splitting at short-circuit. It thus reflects the recombination dynamics and can serve as an indicator of the dominant recombination mechanism [42]. This prior physical knowledge motivates modification of the fitting objective to $J_{shifted}$ instead of the raw $J$. Moreover, this approach allows separate treatment of $J_{sc}$, which is beneficial when calibration errors exist. Such errors, highlighted in discrepancies between experimentally measured values and values calculated from external quantum efficiency EQE and light source spectra, can occur across different device compositions and light sources [44,45] and may be inconsistent. Furthermore, since the NN model is trained with a specific light spectrum, it is impractical to reconstruct the model each time a new device or illumination condition is introduced. As a result, this inconsistency can hinder validation of the estimated material parameters and limit the predictive capability of the model. In this regard, we could potentially resolve the EQE mismatch by adopting a sigmoid-like fitting function for $J_{sc}$.

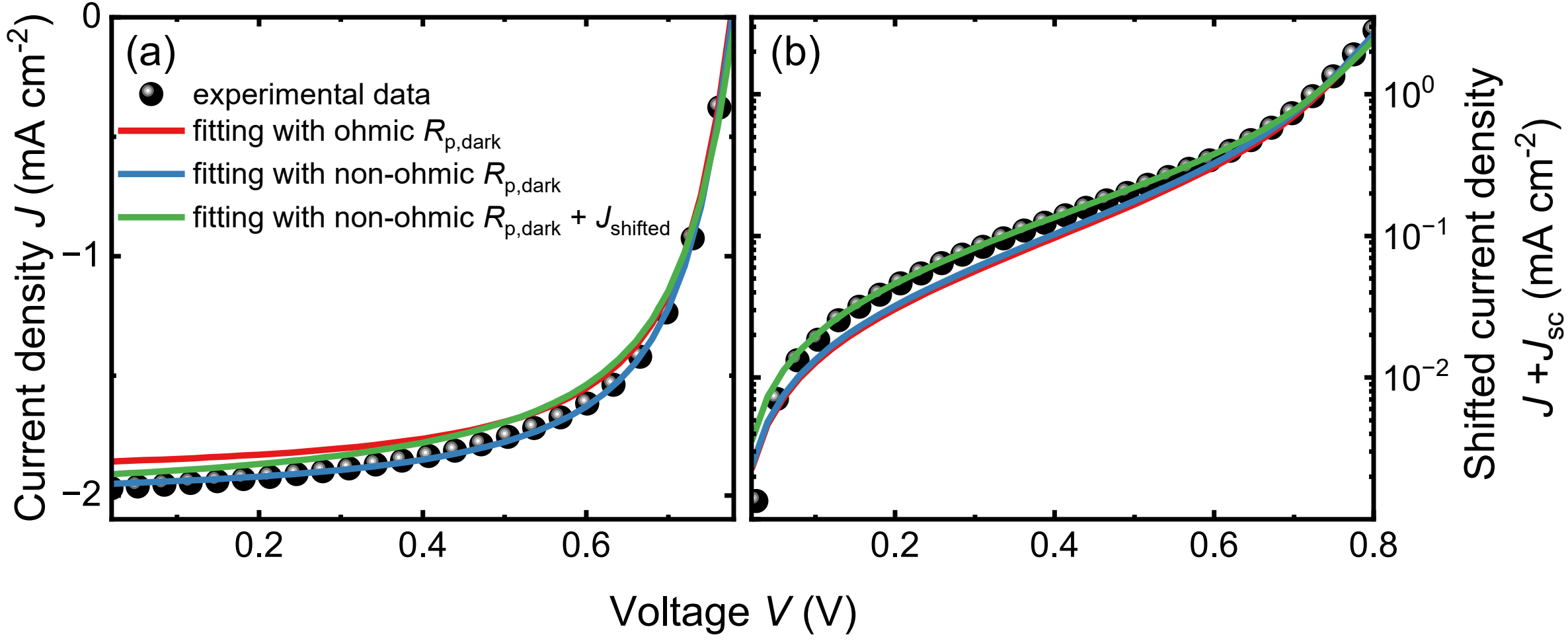

FIG 3. Current-voltage ($JV$) curves of experimental data used for fitting (dots) and the best-fit $JV$ curves obtained under three types of conditions (solid lines) at input light power density $P_{in}$ of 5.5 mW cm$^{-2}$. The fitting using a pre-trained neural network model and CMA-ES genetic algorithm was conducted under the following conditions - assuming an ohmic $R_{p,dark}$ (red), a non-ohmic $R_{p,dark}$ (blue), and a non-ohmic $R_{p,dark}$ with increased weighting on the shifted $JV$ ($J_{shifted} = J + J_{sc}$) compared to the short-circuit current $J_{sc}$ (green).

Based on the prior findings on light-intensity-dependent $JV$ characteristics, we introduce two adjustments to the parameter estimation model. Firstly, we replace the conventional ohmic description of the dark shunt with a non-ohmic model. Second suggestion is to reformulate the fitting objective with the shifted current density to emphasize the recombination dynamics. In the following section, we demonstrate the impact of these modifications by comparing the predictive performance of the inferred parameters.

### B. Influence of physical considerations on parameter estimation

The $JV$ characteristics at light intensity of 5.5 and 72 mW cm$^{-2}$ were simultaneously fitted using a pre-trained NN model and the CMA-ES genetic algorithm, followed by the identification of best-fit parameters $\theta_{best\text{-}fit}$. Detailed information on the construction of the NN model can be found in Supporting Material Note 2 and the corresponding source code is available in the GitHub repository [46]. The material system discussed in the main text is T1:BTP-4F-12, with fabrication details provided in Ref. [27]. Figure 3 illustrates a subset of the best-fit $JV$ curves from the NN model, compared with the input experimental data.

The NN model was originally constructed assuming an infinite dark shunt resistance ($R_{p,dark}$ = $10^{20}$ Ωm$^2$). Therefore, the external $R_{p,dark}$ – either ohmic or nonohmic – can be effortlessly incorporated during the fitting process without retraining the model. The definition of nonohmic shunt resistance in the equivalent circuit model is provided in Supporting Material Note 1. As shown in Figure 3(a) (red and blue solid lines), including the non-ohmic $R_{p,dark}$ results in better agreement with the experimental input, while the extracted $\theta_{best\text{-}fit}$ values are almost identical (TABLE S5 in Supplementary Material [47]).

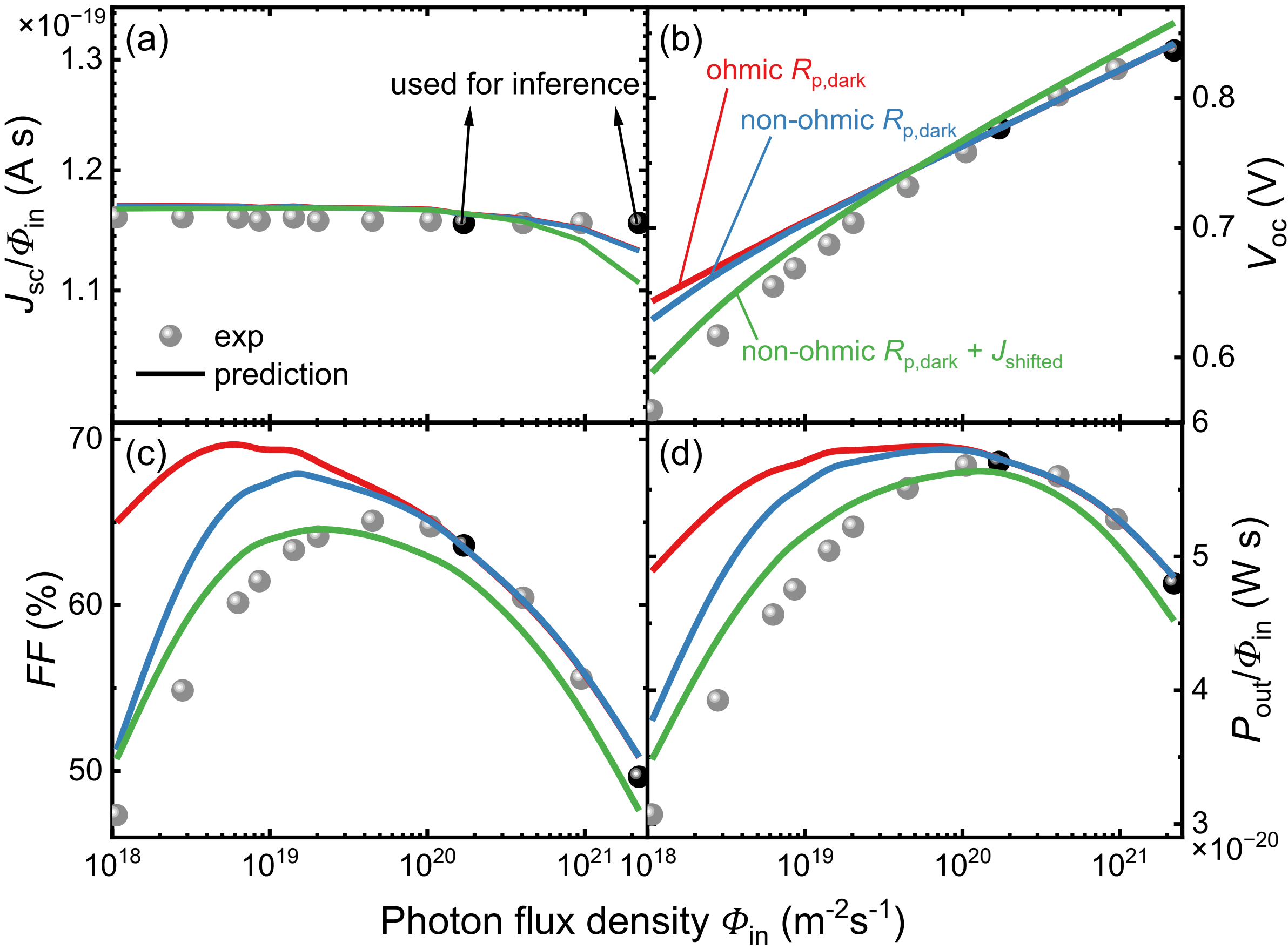

FIG 4. Solar cell parameters of illumination-dependent current-voltage ($JV$) characteristics obtained from PBDB-TF-T1:BTP-4F-12 solar cells (grey dots) and predicted $JV$ characteristics with the extracted best-fit parameters under three different fitting conditions (solid lines): (a) short-circuit current density normalized by the input photon flux density $J_{sc}/\Phi_{in}$, (b) open-circuit voltage $V_{oc}$, (c) fill factor $FF$, and (d) output power density normalized by input photon flux density $P_{out}/\Phi_{in}$. The experimental data used for parameter inference is highlighted in black. The fitting was conducted under the following conditions - assuming an ohmic $R_{p,dark}$ (red), a non-ohmic $R_{p,dark}$ (blue), and a non-ohmic $R_{p,dark}$ with increased weighting on the shifted $JV$ ($J_{shifted} = J + J_{sc}$) compared to the short-circuit current $J_{sc}$ (green).

However, when visualized as $J_{shifted}$ on a logarithmic scale as shown in Figure 3(b), there is deviation from the experimental data even when non-ohmic $R_{p,dark}$ is incorporated. This observation highlights that the standard $JV$ fitting where the error is estimated in the format of raw $J$ often overlooks the low-voltage region, leading to inaccurate estimation of $R_{p,photo}$. To address this, we modified the fitting approach by directly fitting $J_{shifted}$ and separately fitting the $J_{sc}$ with reduced weight (Equation S3 in Supplementary Material [47]). That is, we effectively re-weight the error function to better account for the low-voltage region, instead of arbitrarily designing it. As expected, this strategy resulted in an excellent fit to the shifted $JV$ curves (green solid line), while there is a larger error at the original scale.

Previously, we demonstrated that two modifications in the parameter estimation model result in different fitting results in the format of raw $JV$ and shifted $JV$. Now we compare the predictive performance of the inferred material parameters $\theta_{best\text{-}fit}$, using the experimental data as a baseline. The experimental data were obtained under white LED illumination by adjusting optical density filter and LED current. Figure 4 presents the experimental and predicted $JV$ characteristics plotted against the incident photon flux density $\Phi_{in}$. For better visualization, the short-circuit current density $J_{sc}$ [Fig. 4(a)] and the output power density $P_{out}$ [Fig. 4(d)] are normalized by photon flux density. For reference, the total photon flux density of the standard AM1.5G solar spectrum is approximately $4.3\times10^{21}$ m$^{-2}$s$^{-1}$. While the predicted $J_{sc}$ consistently deviates from the experimental data, partly due to changes in the optical filter during LED measurements [29]

, the prediction accuracy for $FF$ in the low-light regime is increased when a nonohmic $R_{p,dark}$ is incorporated. This highlights the importance of dark $JV$ data in understanding the device mechanism in the low light intensity regime. However, discrepancies between the predicted and experimental data remain under low and intermediate illumination levels. When the shifted *JV* is incorporated in the fitting objective, predictions across light intensities capture the device behavior more completely, though they exhibit reduced predictive power in $J_{sc}$. This observation suggests that the shifted *JV* fitting is more sensitive to inferring $R_{p,photo}$, which may be overlooked in raw *JV* fitting. This discrepancy is more pronounced in the shifted *JV* plot in Figure S2b in Supplementary Material [47] that the predicted shifted *JV*s fail to reproduce the experimental $R_{p,photo}$. In the case of shifted *JV* fitting, best-fit results for $\mu$, $N_{dt}$, and $k_{dir}$ were identified differently compared to raw *J* fitting objective (TABLE S5 in Supplementary Material [47]). Particularly, a higher $N_{dt}$ value of $2.9 \times 10^{15}$ cm$^{-3}$ compared to $4.18 \times 10^{14}$ cm$^{-3}$ obtained from the fitting with *JV* at the original scale emphasizes the influence of defect-assisted recombination in the low- and intermediate-light-intensity regime, bringing into closer agreement with experimental *FF* and $V_{oc}$. On the other hand, the device performance predictions at higher light intensities deviate from the experimental data, likely due to the influence of dark series resistance, which is not fully accounted for in the current workflow.

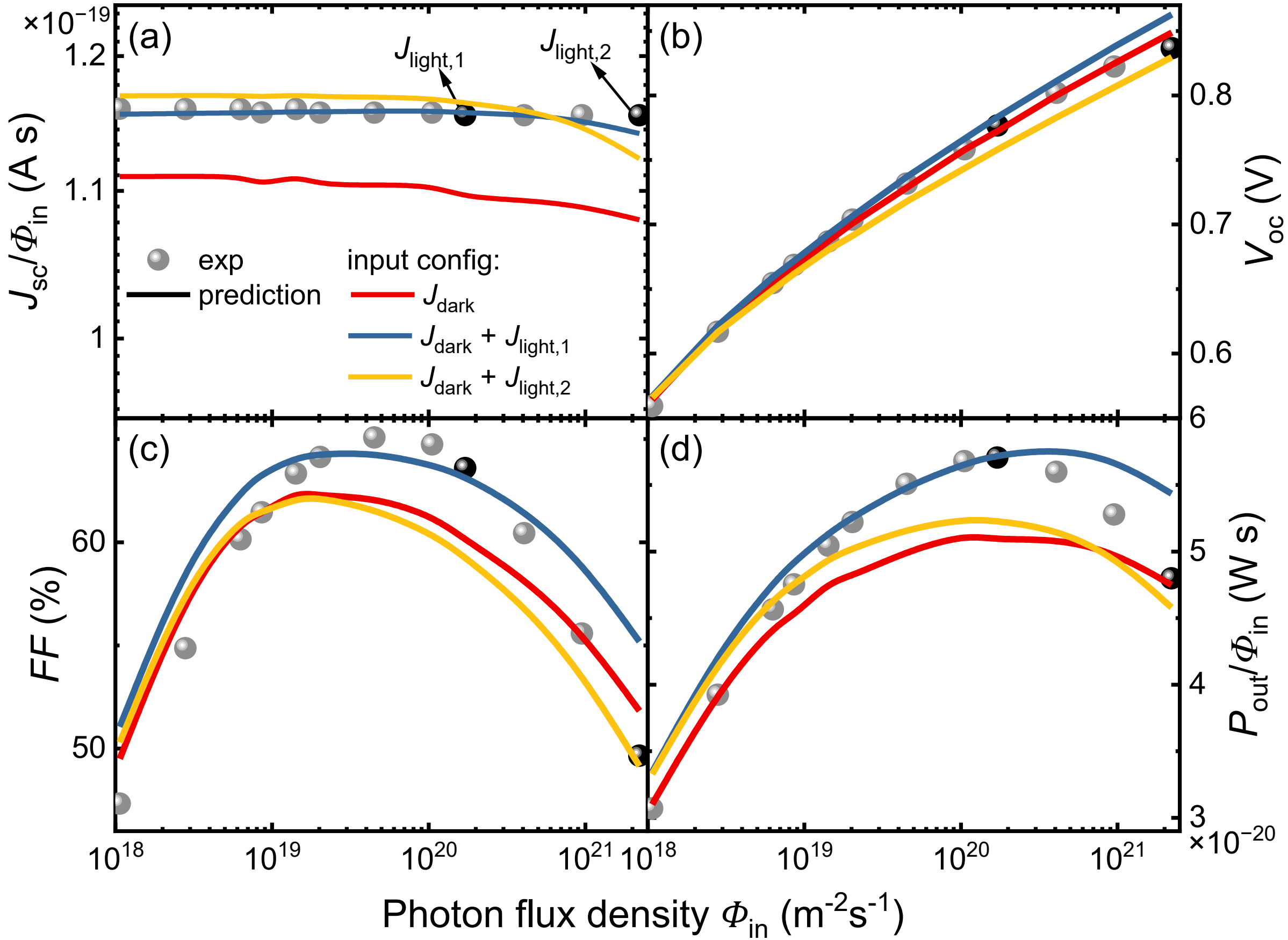

FIG 5. Solar cell parameters of illumination-dependent current-voltage (*JV)* characteristics obtained from PBDB-TF-T1:BTP-4F-12 solar cells (grey dots) and predicted *JV* characteristics with the extracted best-fit parameters from various combinations of input dataset (solid lines): (a) short-circuit current density normalized by photon flux density $J_{sc}/\Phi_{in}$, (b) open-circuit voltage $V_{oc}$, (c) fill factor *FF*, and (d) output power density divided with photon flux density $P_{out}/\Phi_{in}$. The illuminated *JV* curves used for Bayesian inference are highlighted in black color and labelled as $J_{light,1}$ and $J_{light,2}$.

### C. Influence of input data configuration on parameter estimation

In this last section, we investigate how the predictive power depends on the different configuration of the input dataset. In the previous section, it is demonstrated that dark *JV* contains significant information for parameter estimation. Given the dark *JV* as a baseline, additional *JV* curve either at 5.5 mW cm$^{-2}$ (denoted as $J_{light,1}$) or 72 mW cm$^{-2}$ ($J_{light,2}$), is incorporated in the parameter estimation workflow, and we evaluate how well the inferred parameters capture the light-intensity-dependent *JV* characteristics. Subsequently, the prediction results are shown in Figure 5.

Interestingly, using only the dark $JV$ curve for inference results in reasonably accurate predictions of $V_{oc}$, while the $FF$ is not well captured, particularly under intermediate illumination levels. Once an illuminated $JV$ curve $J_{light,1}$ is incorporated, the $FF$ predictions are more consistent with experimental data. However, when the light $JV$ at high intensity ($J_{light,2}$) is used in combination with the dark $JV$, the prediction accuracy decreases. This discrepancy can be attributed to the increasing influence of series resistance at higher light intensities, which is not included in NN model. Consequently, the recombination losses that should be ascribed to the series resistance are instead misattributed to a higher direct recombination coefficient (Table S6 in Supplementary Material [47]), ultimately degrading the accuracy of light-intensity-dependent performance predictions. On the other hand, poor prediction accuracy using only the dark $JV$ curve likely stems from its limited information content. For example, exciton dissociation probability, which eventually determines the photogenerated charge carrier density, cannot be extracted from dark $JV$ data alone. While $J_{dark}$ contains information about recombination dynamics such as saturation current, it lacks insights into how the quasi-Fermi-level splitting evolves under illumination.

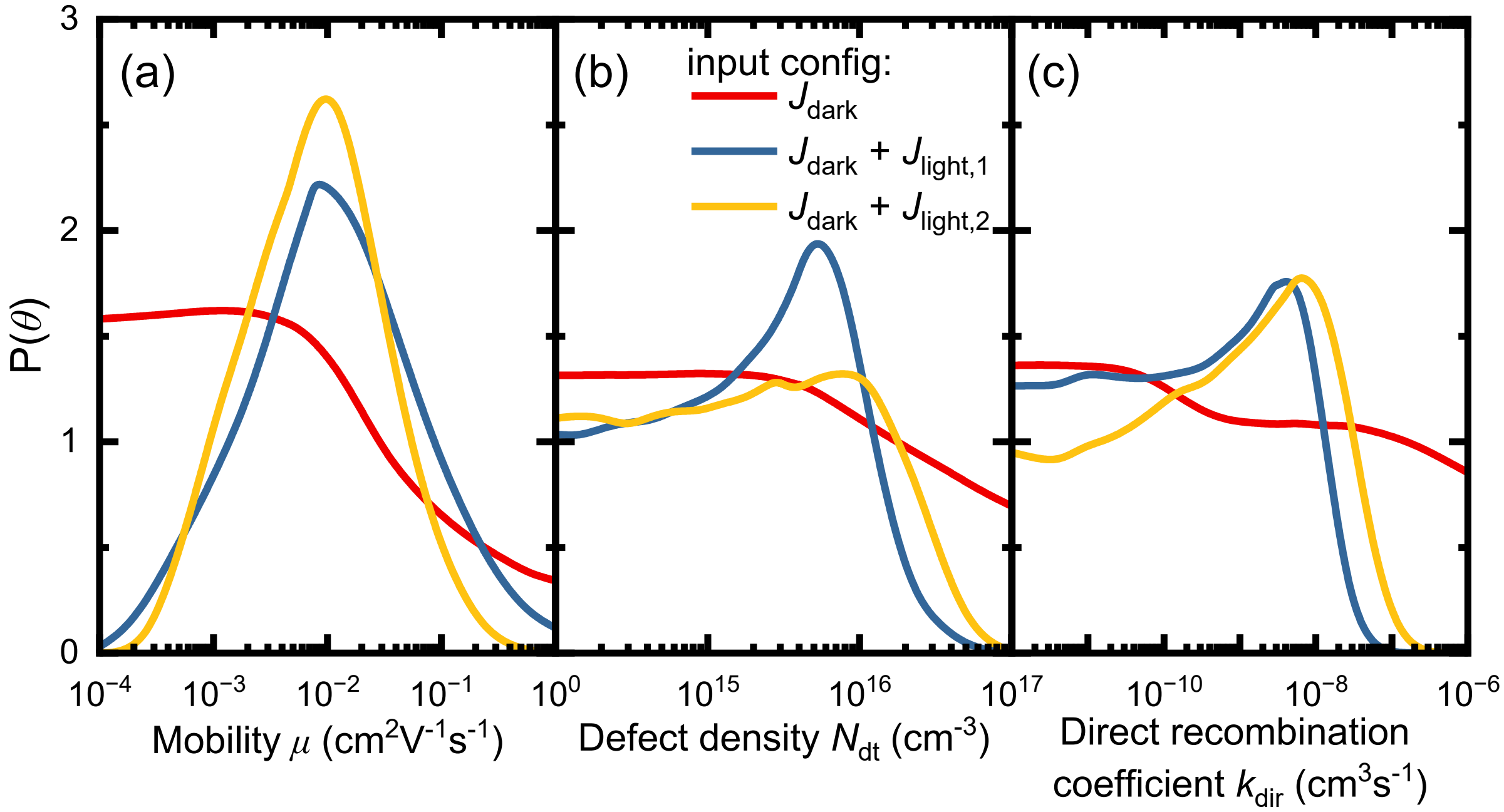

FIG 6. Projected multivariate probability densities onto one-dimensional space for (a) mobility $\mu$, (b) defect density $N_{dt}$, and (c) direct recombination coefficient $k_{dir}$ obtained from parameter estimation using three different types of input dataset – a single dark $JV$ curve (red), one dark and one light $JV$ curve at 5.5 mW/cm² (blue), and one dark and one light $JV$ curve at 72 mW/cm² (yellow). $J_{light,1}$ refers to the $JV$ curve measured at 5.5 and $J_{light,2}$ at 72 mW/cm².

This parameter uncertainty is further visualized in the probability distribution plots for key electronic parameters in Figure 6. The probability densities for $\mu$, $N_{dt}$, and $k_{dir}$ were determined using Equation S5 in Supplementary Material [47] and subsequently the posterior probabilities were updated upon additional light $JV$ input based on Bayes' rule. When a single $J_{dark}$ is used, the distributions are relatively uniform and have no clear peaks, indicating high uncertainty. In contrast, the inclusion of a light $JV$ curve produces distinct peaks in the distributions, enabling more reliable parameter estimation. This observation supports the limited information content in dark $JV$ data. Consequently, the input configuration of $J_{dark}$ and $J_{light}$ preferably at a light intensity where the influence of series resistance is not prevailing would be sufficient to accurately capture the light-intensity-dependent behavior of solar cells.

## III. CONCLUSIONS

While the machine learning techniques enable efficient estimation of material parameters using a neural network (NN) surrogate mode, the reliability of the inferred parameters depends on hidden assumptions such as the reduction of the parameter space and the underlying physical representations embedded in the NN model. Since multiple parameter combinations can often reproduce the identical experimental observation, it is difficult to validate the correctness of the inferred parameters from the fit alone. We therefore used the criterium *predictive* *power* for evaluating a parameter estimation model. Predictive power serves as a useful metric since it quantifies whether the inferred parameter combination can reproduce device behavior under illuminations beyond the point used for fitting. This constitutes 'usefulness' according to what could be a sensible definition of the term 'useful', because the purpose of having a model and inferred model parameters is generally to be able to understand and predict the performance of a device outside the range of measured parameters. This is akin to the training of a neural network whose usefulness only becomes apparent, when it is tested against the test dataset that was not used for training.

In this manuscript, we applied the criterion of predictiveness to light-intensity-dependent *JV* characteristics of organic solar cells as a case study. Within this case study, we varied two aspects of the estimation model – the treatment of dark shunt resistance and the choice of fitting objective (raw $J$ vs. shifted current $J + J_{\mathrm{sc}}$), which are factors considered to be relevant for illumination-dependent solar cell performance. Our analysis revealed that predictive performance strongly depends on the choice of parameter estimation model and these variations are consistent with established physical principles. Incorporating a non-ohmic $R_{\mathrm{p,dark}}$ brought predicted $V_{\mathrm{oc}}$ and $FF$ into closer agreement with experimental results particularly at low light intensities, consistent with shunt leakage being the dominant loss mechanism in that regime and with the ohmic description being unable to represent it. On the other hand, fitting on raw $J$ systematically underweighted the low-voltage region, which limited the accurate inference of the photoshunt behavior. Consequently, reformulating the fitting objective toward shifted current densities resolved the information at low voltages and was able to capture light-intensity-dependent recombination dynamics more completely at the cost of a larger deviation in predicted $J_{\mathrm{sc}}$. We emphasize here that neither estimation model was objectively "better", each put an emphasis on a different region, and the assessment of predictive performance is what made that difference visible rather than it being hidden inside an equally-good fit.

Finally, we evaluated predictive power for different input dataset configurations and observed that performance can be limited by the information content of the data used for inference. Using only a dark *JV* curve gave reasonable but incomplete predictions, since dark data contains no information on photogenerated charge carrier behavior. Incorporating at least one illuminated *JV* consistently improved prediction accuracy, while a curve measured at high light intensity ($\geq 0.1$ suns) introduced series resistance effects that were not accounted for in the current workflow. Together, these results establish predictive performance as a general framework for evaluating parameter estimation models.

## ACKNOWLEDGMENTS

The authors acknowledge funding by the Helmholtz Association via the POF IV funding as well as via the SolarTap project. The authors further thank the Ministry of Economic Affairs, Industry, Climate Action and Energy of the State of North Rhine-Westphalia for funding via the ENFA project.

## DATA AVAILABILITY

The data of the figures in this article are openly available via 10.5281/zenodo.21435795 and PYTHON scripts used for training neural network models and fitting the data are available from [46].